\documentclass[fleqn,twoside]{article}
\usepackage{espcrc2}

\usepackage{epsfig}

\newcommand{\AmS}{{\protect\the\textfont2
  A\kern-.1667em\lower.5ex\hbox{M}\kern-.125emS}}

\title{The RICH counter of the AMS experiment 
\footnote{Talk given at the RICH2002 conference, Pylos (Greece), June 8-12, 2002}
}

\author{M. Bu\'enerd \address[MCSD]{Institut des Sciences Nucl\'{e}aires, IN2P3/CNRS, 
53 av. des Martyrs, 38026 Grenoble cedex, France}
        \thanks{For the AMS RICH collaboration: Bologna INFN, Grenoble ISN, Lisbon LIP,
Madrid CIEMAT, Maryland U., and Mexico UNAM }
}

\begin{document}

\begin{abstract}
The RICH counter of the AMS experiment is described and its 
expected performances discussed. Prototype results are reported.
\vspace{1pc}
\end{abstract}

\maketitle


\section {INTRODUCTION}
%
\begin{figure}[htb]
\begin{center}
\vspace{-2cm}
\epsfysize=10cm
\epsfbox{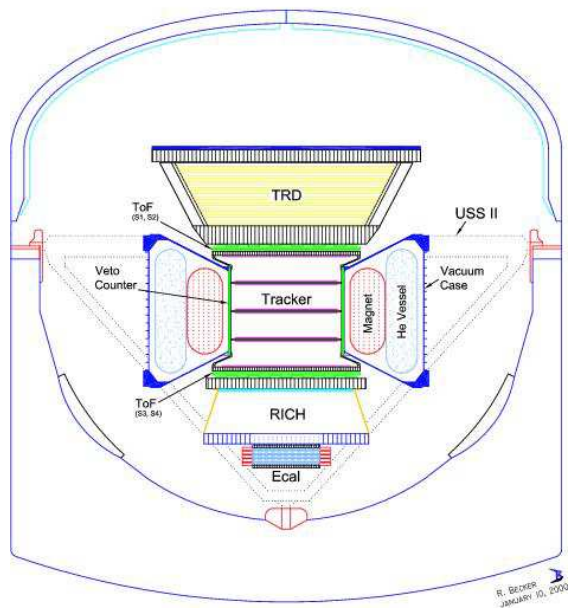}
\end{center}
\vspace{-0.5cm} 
\caption{\it Schematic view of the AMS02 spectrometer architecture on its 
Unic Support Structure (USS) in the Space Shuttle bay.
}
\label{AMS2}
\end{figure}
The second phase of the AMS experiment should begin on the International Space Station in the 
year 2005. In this contribution, the Cherenkov imager (RICH) of the AMS spectrometer, its 
Physics motivations and its main features and expected performances are reviewed. 

The particle spectrometer shown on fig~\ref{AMS2} will be able to accumulate statistics larger 
by 3 to 4 orders of magnitudes than those measured so far by other embarked experiments, for all 
the species of particles studied. These capabilities will allow to address with an unmatched 
sensitivity the main scientific objectives of the program: 1) The search for primordial 
antimatter in space (anti$^4He$ and anti$^{12}C$ nuclei); 2) The search for dark matter in space 
through the signature of neutralino annihilations in the $\bar{p}$ and e$^+$ spectra.
In addition, this search will also allow to achieve a high statistics study of many species 
of the cosmic ray population, including $e^+$, $e^-$, $p$, $\bar{p}$, and the lightest nuclei 
isotopes, $d,t,^{3,4}$He. 
Heavier isotopes will also be studied over the range of mass and charge identification of 
the RICH as discussed below. Unstable ions with long lifetime like $^{10}$Be, and $^{26}$Al 
are of particular interest since they provide a measurement of the time of confinement of 
charged particles in the galaxy (galactic chronometers). 

\begin{figure}[htb]
\begin{center}
\epsfysize=7cm
\epsfbox{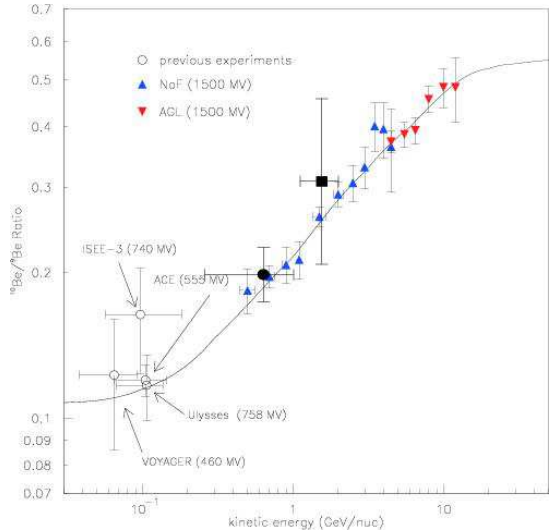} 
\end{center}
\vspace{-0.5cm}
\caption{\it Expected statistics for the $^{10}Be$ isotope for 6 weeks of counting
from a simulation using the RICH of AMS assuming the use of two radiators: NaF and aerogel with
n=1.03. The two large full square and full circle symbols correspond to the recent measurements 
of the ISOMAX experiment \cite{RBE}. See text and \cite{BO00} for the other refs.
}
\label{BE10}
\end{figure}

This is illustrated on figure~\ref{BE10} with the simulation result for $^{10}$Be  \cite{BO00}.
Six weeks of counting would provide a highly accurate data sample over a largely unexplored 
range of momentum.
%
\section{THE RICH COUNTER}
The RICH counter will allow the measurement of the mass of isotopes ($A$) and of the charge of 
elements ($Z$) up to maximum values of the order of 20 at best for both $A$ and $Z$, depending 
on the final configuration of the counter. The momentum range for mass separation of isotopes 
should be 1-12~ GeV/c, while for the charge it should extend over the whole momentum dynamics 
of the spectrometer.   
The RICH will also contribute to the $e^-/\bar{p}$ and $e^+/p$ discrimination and to the Albedo 
particle rejection. 
%
\begin{figure}[htb]
\begin{center}
\epsfysize=9cm
\epsfbox{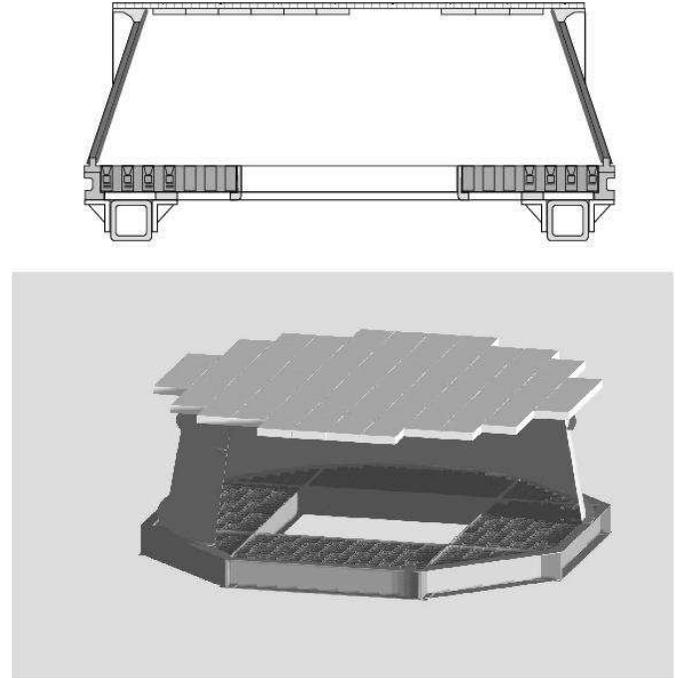} 
\end{center}
\vspace{-1cm} 
\caption{\small\it   Side cut view (top) and Cut view in perspective (bottom) of the RICH 
counter showing the radiator plane at the top separated from the photodetector plane by the 
(photon) drift space (ring expansion gap). The conical mirror encloses the drift space.
}
\label{RICH}
\end{figure}

\subsection{DESIGN}
See refs \cite{SIMU} and \cite{PROTO1} for a general discussion of the topic.
The general principle of the counter and its design had to comply with several types of drastic 
limitations specific to embarked experiments, on the volume, weight (currently about 190~kg) 
and electric power consumption of the counter (currently about 150~W), and with the long term 
reliability requirement of the instrument and of its components. It had also to be compatible 
with the stray magnetic field of the superconducting magnet, which will reach close to 300~G in 
the photodetector volume. The proximity focussing principle, using solid state radiators and 
photomultiplier (PMT) detectors, has been considered as the most suitable technique to meet all 
the above requirements. Fig~\ref{RICH} shows a schematic view of the design. The radiator plane 
at the top is separated from the photodetector plane by a 45~cm (photon) drift space (ring 
expansion gap). The empty space in the detector plane corresponds to the location of the 
electromagnetic calorimeter.  A conical mirror encloses the drift volume to increase the 
acceptance. 
\par\noindent
\begin{figure}[htb]
\begin{center}
\epsfysize=6cm 
\epsfbox{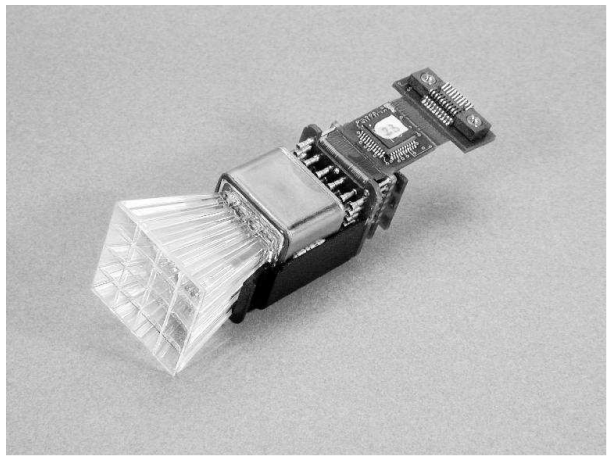} 
\end{center}
\vspace{-0.5cm} 
\caption{\it Detector cell including PMT, front end electronics, light guide matrix and (half) 
housing shell. The cell fits inside a shielding tube (not shown).  
}
\label{CELL}
\end{figure}
%
\begin{figure}[htb]
\begin{center}
\hspace{1cm}
\epsfysize=10cm
\epsfbox{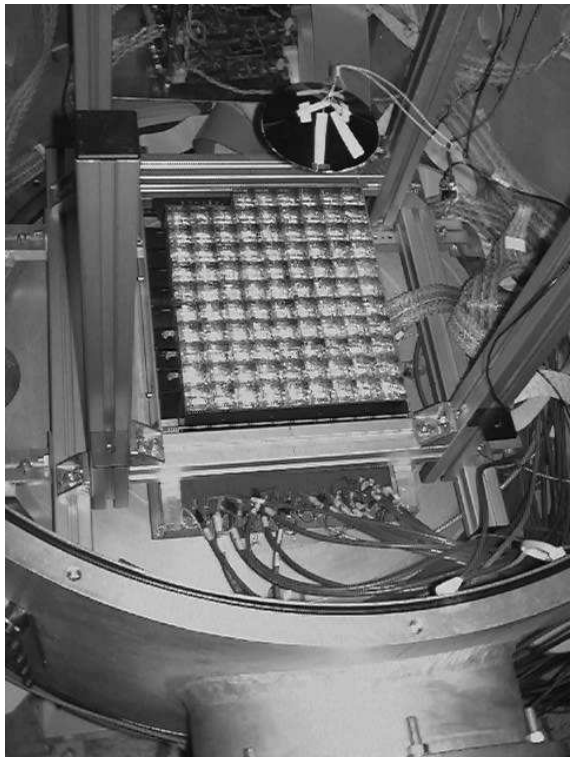}
\end{center}
\vspace{-0.5cm} 
\caption{\it Matrix of the 96 detector cells of the second generation prototype in its 
testing environment.
}
\label{MATR}
\end{figure}
$\bullet\;${\bf Mechanical structure of the photodetector plane :}
The detector support is based on a grid structure providing the mechanical stiffness, in which 
the individual detector modules and their magnetic shielding envelopes are lodged \cite{GAVA}. \\
$\bullet\;${\bf Photodetectors :}
The photomultiplier selected is the 16-anode R7900\_M16 from Hamamatsu \cite{HAMA}, which
individual anode size is, although not the optimum value, the smallest compatible with a realistic 
design. The chromatic range is limited at short wave lengths by the Borosilicate window cutoff 
of the PMTs. \\
$\bullet\;${\bf Front end electronics :}
It is based on a spectroscopy chain involving a charge 
preamplifier. A track-and-hold system allows the 16 channels of the PMT to be multiplexed, encoded
in sequence, and read by the DAQ system \cite{FEISN}. \\
$\bullet\;${\bf Photodetector modules :} 
A module includes a matrix of light guides coupled to a 
PMT, connected to its socket and front end electronics readout. These elements are enclosed in 
a plastic shell as shown on fig~\ref{CELL}. \\
$\bullet\;${\bf Radiators :}
The final choice for the radiator hasn't been fixed yet. However in the currently considered 
solution two radiators would be used: A small patch, $\approx$~25x25~cm$²$ of Sodium fluoride 
(NaF, 5~mm thick) in the central region of the radiator area, and Silica aerogel with index 
n=1.03, 30~mm thick, over the rest of the area, would allow to cover the range in momentum 
between about 1 and 12~ GeV/c per nucleon. \\
%
\section{PROTOTYPES}
%
\begin{figure}[htb]
\begin{center}
\epsfysize=8cm
\epsfbox{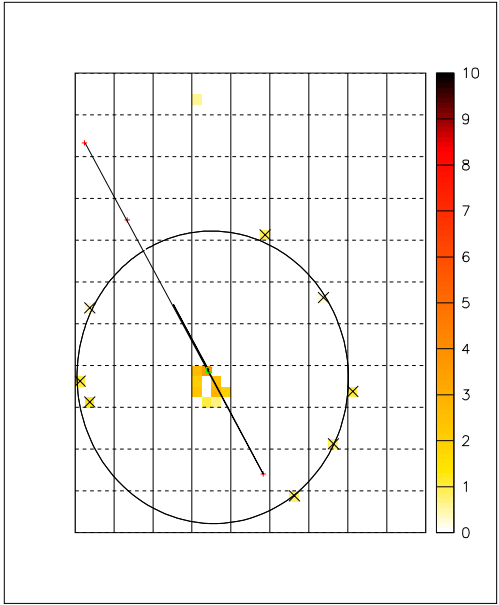} 
\end{center}
\vspace{-1cm} 
\caption{\it Typical CR event measured with the second generation prototype. The figure shows the 
photon hit positions and the fit Cherenkov ring. The hit cluster at the center of the ring 
corresponds to the particle hit on the matrix of light guides. The crosses along the straight line
are the projections on the detector plane of the particle hits on the tracker planes placed above 
and below the detector, with a linear regression fit.
}
\label{EVT}
\end{figure}
%
\begin{figure}[htb]
\begin{center}
\epsfysize=6cm
\epsfbox{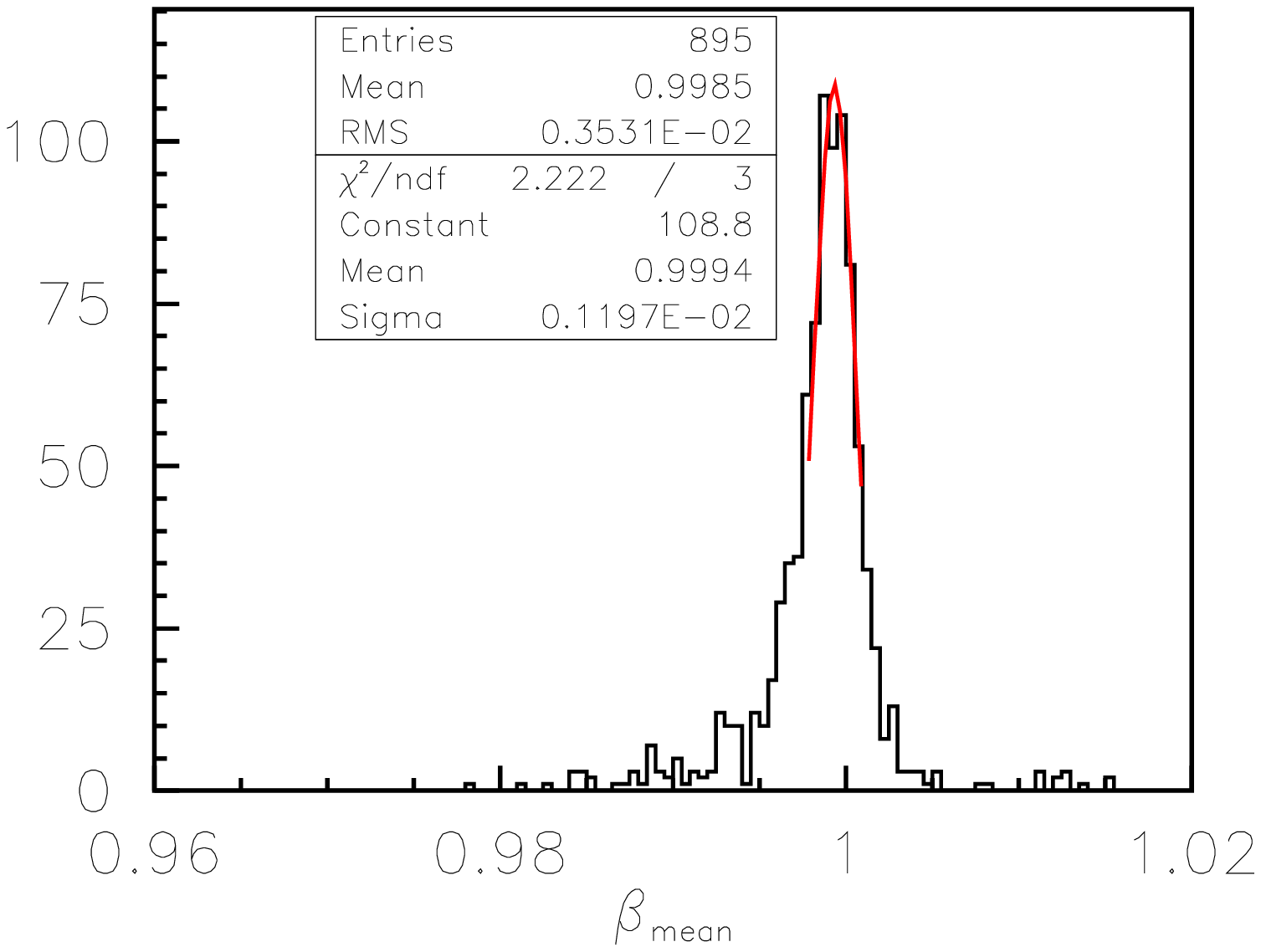}
\epsfysize=6cm
\epsfbox{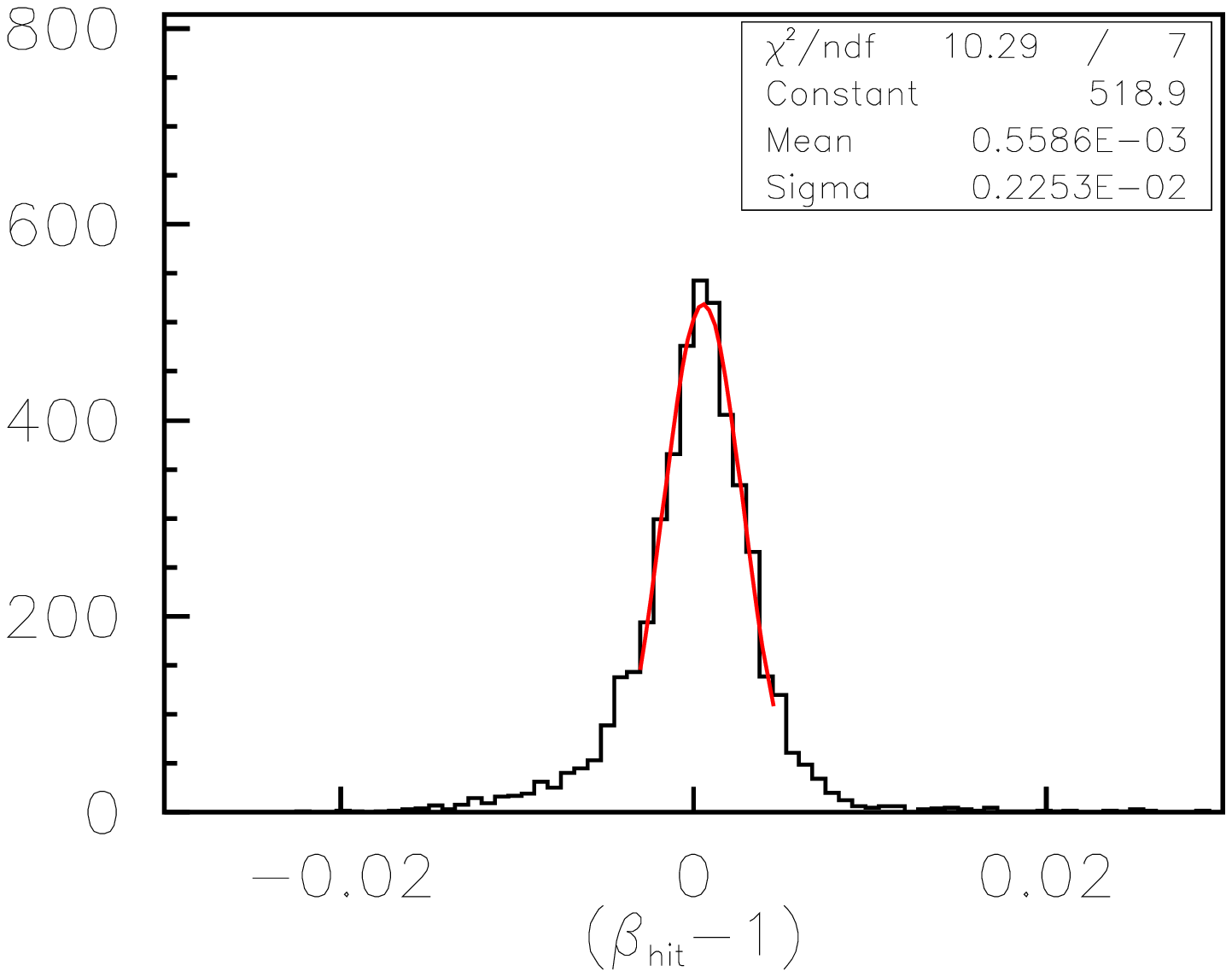} 
\end{center}
\vspace{-1cm} 
\caption{\it Mean (top) and single hit (bottom) velocity distributions measured for a sample of CR 
particles. See text for details.
}
\label{RESOL}
\end{figure}
Prototypes have been built to investigate the performances of the counter for velocity and charge 
measurement. The velocity resolution results from contributions arising from radiator chromatism, 
radiator thickness, and ratio of photodetector pixel size over drift distance, which have to be 
balanced for an optimized design (see \cite{SIMU} for details).
The study of a first generation prototype has been completed two years ago \cite{PROTO1} and the 
second generation which incorporates most of the final detector elements, is currently being tested.
Both counters were operated in a setup including 2 plastic scintillator paddles to define the 
geometrical acceptance and for trigger definition, and a simple tracker based on a set of 3 $xy$ 
mwpcs \cite{PROTO1}.

The results with Cosmic Ray particles are illustrated on figs~\ref{EVT} and \ref{RESOL}. The 
former shows a typical event obtained with an aerogel radiator of refractive index 1.03 (SP30 from 
Matsushita \cite{MATSU}). The hit cluster at the center of the ring corresponds to the particle 
impact on the light guide + PMT system, which provides a useful complementary information on the 
particle ID and trajectory. 
Fig \ref{RESOL} shows the distribution of the reconstructed $\beta$ for a sample of CR particles 
(mainly muons). The velocity resolution is found to be of the order of 
$\frac{d\beta}{\beta}\approx 2.2 10^{-3}$ per photon ($1.2 10^{-3}$ per event), i.e., close to the 
limit set by the three dominant contributions to the resolution for this material: 0.85 (chromatic 
dispersion), 1.1 (pixel size/drift gap), 1.15 (radiator thickness), in units $10^{-3}$, combining to 
about 2 per photon. 

The results are significantly less good for the SP50 aerogel with refractive index 1.05, in account 
of both the larger chromatic dispersion and significantly less good clarity coefficient of the 
material.

With a NaF radiator, the resolution obtained with Cosmic Ray muons is of the order of $1.610^{-2}$ 
per photon ($1.15 10^{-3}$ per event) with a short drift gap of 7.5~cm
More details on the analysis are given in F.~Barao's presentation at this conference \cite{FBAR}. 

The response of the counter to ions will be studied to evaluate accurately its performances for 
charge measurement, by using a dedicated test beam obtained from the CERN SPS ion beam colliding on 
a fragmentation target. The produced fragments will be magnetically selected in rigidity by the beam 
magnetic analyzer, over the range of mass of interest for the counter, i.e., from charge 1 (Hydrogen) 
to about 26 (Fe region) \cite{TESTB}.
\section{SUMMARY}
In summary, it has been shown that the design of the RICH counter of the AMS experiment is now 
completed. The instrumental and technical solutions have been successfully tested over two 
generations of prototypes. The construction of the flight model of the counter will start on 
january 2003.

{\sl acknowledgements}: The author is indebted to his collaborators B. Baret and L. Derome for their 
help in the preparation of this talk.  


\end{document}